\documentclass[aps,prl,preprint]{revtex4}
\usepackage{epsfig}
\begin{document}

\title{\Large\bf Anisotropy and asymmetry in fully 
developed turbulence}
\author{\normalsize {S. I. Vainshtein}\\
{\small\it Department of Astronomy and Astrophysics, University of Chicago
}}

\begin{abstract}
Using experimental longitudinal and transverse velocities data for very high Reynolds number
turbulence, we study both anisotropy and asymmetry of turbulence.
These both seem to be related to small scale turbulent structures, and  to
intermittency. We may assume that the large scale velocity shear gives an impact into
the small scale turbulence, resulting in non-locality, and related anomalous events.
\end{abstract}

\maketitle

\section{Introduction}
Recently considerable attention has been devoted to the study of local isotropy 
of the 
high Reynolds number turbulence suggested by \cite{k41}. The turbulence is stirred at 
the large scales, and this energy is transported into the small scales, where, after
intense nonlinear interaction, it becomes isotropic. Deviation from isotropy would
mean that there is direct interaction between large scales containing non-universal anisotropy
and small scales, leading to non-universal behavior of small scale spectral properties.

It was indeed shown experimentally that in a  sheared turbulence the 
isotropy is not 
sufficiently restored for both scalar and velocity fields, see \cite{001}, \cite{02},  \cite{1r}, 
\cite{2r}, \cite{new2}, \cite{03}, \cite{new3}. 

High Reynolds number (atmospheric) turbulence also shows deviations from isotropy. It was
shown that the large scale shear does contribute into the scaling of the structure
functions, see \cite{new1}. Anisotropic scaling of high-order structure functions was studied
by \cite{Susan}. It was shown that the anisotropy in small scales
remains stronger than expected before. The SO(3) decomposition was used to describe the
anisotropy, see \cite{Susan}, \cite{Biferale1}, \cite{Biferale3}, \cite{Biferale4}. The persistent
anisotropy in small scale turbulence was found to be related to the intermittency corrections, 
\cite{Biferale2}.

In numerical simulations, the failure to
return to isotropy was linked to both  asymmetry of the probability distribution function 
(PDF) and to the vortex sheets,  \cite{3r}. It became clear that the shear in the integral
scale induces asymmetry
down to the small scales, where it is manifested by intermittent structures like cliffs, etc., \cite{4r}. 

On the other hand, the asymmetry PDF was found to be related to the intermittency, \cite{94}. 
Thus, these three items, i.e., anisotropy, asymmetry and intermittency seem to be related.
Note, however, that, in principle, these items are independent of each other. 
For example,   the asymmetry of the PDF appears naturally in turbulence, even without any 
anisotropy. 
Denote $u$ - longitudinal and $v$ - transverse (vertical) components of the velocity,
and $u_r=u(x+r)-u(x)$, $v_r=v(x+r)-v(x)$, the velocity increments. 
Then, we have,
\begin{equation}
\langle u_r\rangle=0, ~~~ {\rm and} ~~~ \langle v_r\rangle=0,
\label{first_m}
\end{equation}
and (in inertial range)
\begin{equation}
\langle u_r^3\rangle= -\frac{4}{5}\varepsilon r,
\label{kolm}
\end{equation}
the so-called 4/5-Kolmogorov law \cite{law}. Besides,
\begin{equation}
\langle u_rv_r^2\rangle=\frac{1}{6}\frac{d\langle u_r^3\rangle}{dr},
\label{mixed0}
\end{equation}
\cite{book}, \cite{Monin}. The fact that the first moments vanish, see (\ref{first_m}),
 whereas the two third moments
(\ref{kolm}) and (\ref{mixed0}) do not clearly indicate that both the PDF for $u_r$ and
the joint PDF for $u_r$ and $v_r$ are asymmetric.

As the Kolmogorov law is derived assuming isotropic turbulence, this asymmetry  in principle exists
without anisotropy. Besides, the scaling defined by the Kolmogorov law (\ref{kolm}) does not have any intermittency
corrections. Nevertheless, as mentioned above,
the asymmetry (even without anisotropy) may be related to the intermittency of turbulence,
as suggested by \cite{94}, \cite{96a} and \cite{asym}. Still, this connection does not seem to present 
the whole
picture. Presumably, as in sheared flows, the anisotropy should also be taken into 
considerations.  As mentioned above, even very high-Reynolds-number turbulence manifests
anisotropy in small scales, which is higher than predicted from dimensional arguments by
\cite{Lumley}. This suggests that there is additional to  Kolmogorov cascade transfer of 
energy from large scales directly to small, leading to anomalous events like intermittency and anisotropy.

In this paper we focus on these events. In particular, we construct experimental joint
PDF for $u_r$ and $v_r$ to see directly if anisotropy and asymmetry is present in rare
violent events (responsible for the intermittency). Note that constructing
the joint 2D PDF's proved to be useful in turbulence,   see  for example \cite{joint1},
\cite{joint2}, \cite{joint3}. Even more information about the connection between anisotropy,
asymmetry and intermittency we obtain from several conditional and cumulative averages described
in the next section.

\section{Problem description}

We will work with dimensionless variables,
$$
u'_r=\frac{u_r}{\langle u_r^2\rangle^{1/2}}, ~~~~
v'_r=\frac{v_r}{\langle v_r^2\rangle^{1/2}},
$$
and  construct experimental joint PDF, $p(u'_r,v'_r)$ to study both asymmetry and
anisotropy. This 2D distribution is useful to compare with 2D Gaussian anisotropic
distribution, $p_G$, see Appendix, (\ref{GaussA}), (\ref{Gauss1A}).

The joint 2D PDF gives general information about the distribution. More detailed information
which is easier to analyze are provided by different 1D distributions and cumulative moments.
In particular, we are interested in studding the third mixed moment, (\ref{mixed0}),
\begin{equation}
\langle u_rv_r^2\rangle=\int u_rv_r^2p(u_r,v_r)du_rdv_r,
\label{mixed}
\end{equation}
that is, it is important to find out what part of the distribution contributes most into this moment.
It was shown before, see \cite{jfm}, that the tail parts of 1D distributions satisfactory recover the
moment. 

Additional information is given by conditional average,
\begin{equation}
\langle v_r^2|u_r\rangle=\int v_r^2 p(v_r|u_r) dv_r=\frac{\int v_r^2 p(v_r,u_r)dv_r}{p(u_r)}=
\frac{\Phi(u_r)}{p(u_r)},
\label{mixed_c}
\end{equation}
or, in dimensionless variables,
\begin{equation}
\langle {v'}_r^2|{u'}_r\rangle=\int p({v'}_r|{u'}_r){v'}_r^2 d{v'}_r=\frac{\Phi({u'}_r)}{p({u'}_r)},
\label{mixed_c_d}.
\end{equation}
In these expressions we introduced
\begin{equation}
\Phi(u_r)=\int v_r^2p(u_r,v_r)dv_r, ~~~\Phi({u'}_r)=\int {v'}_r^2p({u'}_r,{v'}_r)d{v'}_r
\label{Phi}
\end{equation}
Therefore,
\begin{equation}
\langle u_rv_r^2\rangle=\int  u_r\langle v_r^2|u_r\rangle  p(u_r)du_r=\int u_r\Phi(u_r)du_r,
\label{mixed1}
\end{equation}
\begin{equation}
\langle {u'}_r{v'}_r^2\rangle=k_a=
\int{u'}_r\langle{v'}_r^2|{u'}_r\rangle p({u'}_r)d{u'}_r=\int {u'}_r\Phi({u'}_r)d{u'}_r
\label{mixed2}
\end{equation}

The function $\Phi(u_r)$ deserves special attention. It follows from (\ref{Phi}) that
\begin{equation}
\int \Phi(u_r)du_r=\langle v_r^2\rangle,
\label{Phi0}
\end{equation}
that is, in a way, the function is a $v_r^2$-distribution versus $u_r$. The first moment 
of this distribution should not vanish, as it follows from (\ref{mixed1}). This means
that the $v_r^2$-distribution should be be asymmetric.
For Gaussian distribution,   the function $\Phi_G$ can be
easily calculated, see Appendix, (\ref{GaussPhiA}).

We used x-wire data  acquired at Brookhaven National Lab.
Distance of probe above the ground: 35m; Number of samples: 40960000 per component, that 
is, for longitudinal ($u$) and transverse ($v$) components. Sampling frequency: 5 kHz. 
Mean velocity: 5.15076224 m/s;
rms $u$-velocity: 1.81617371 m/s;
rms $v$-velocity: 1.3646025 m/s;
Taylor Reynolds number: 10680 (courtesy of Sreenivasan). As usual, the data are interpreted
using the Taylor's hypothesis. 

All throughout the paper we process data  for closest two samples, that is,
$r$ corresponds to the smallest distance between two samples.

\begin{figure}
  \includegraphics[height=8cm,width=16cm]{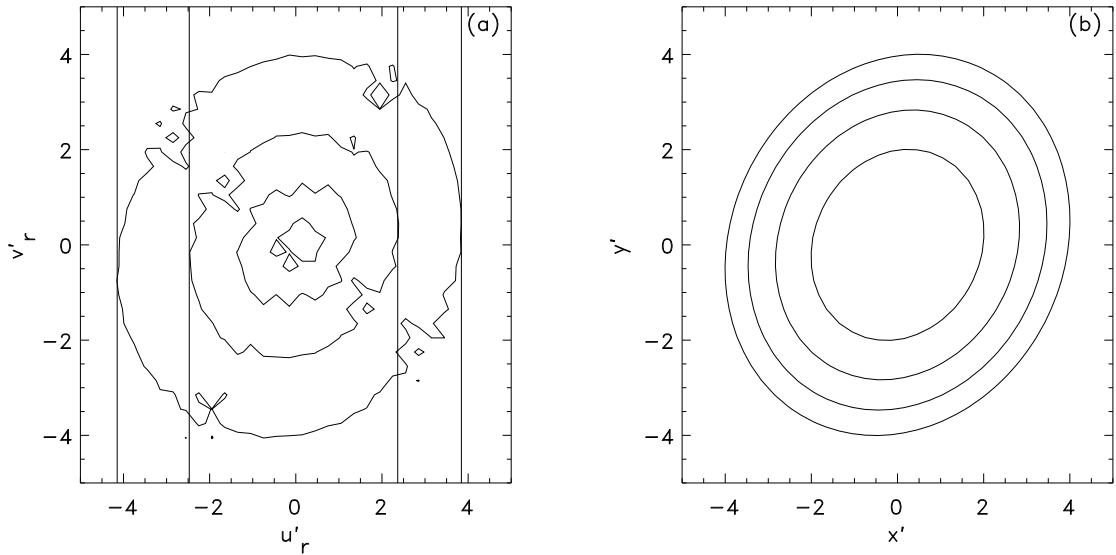}
  \caption{ Experimental joint PDF, core part. The indicated levels correspond to 
$e^{-2},~e^{-4},~e^{-6},~e^{-8}$ of the maximum of the PDF, inside out correspondingly.
The contours a) depict experimental PDF, and b) -- Gaussian anisotropic (\ref{Gauss1A}).
}
\vspace{1cm}
\label{fig1}
\end{figure}
\section{Joint PDF}
\begin{figure}
  \centerline{\includegraphics[height=8cm,width=16cm]{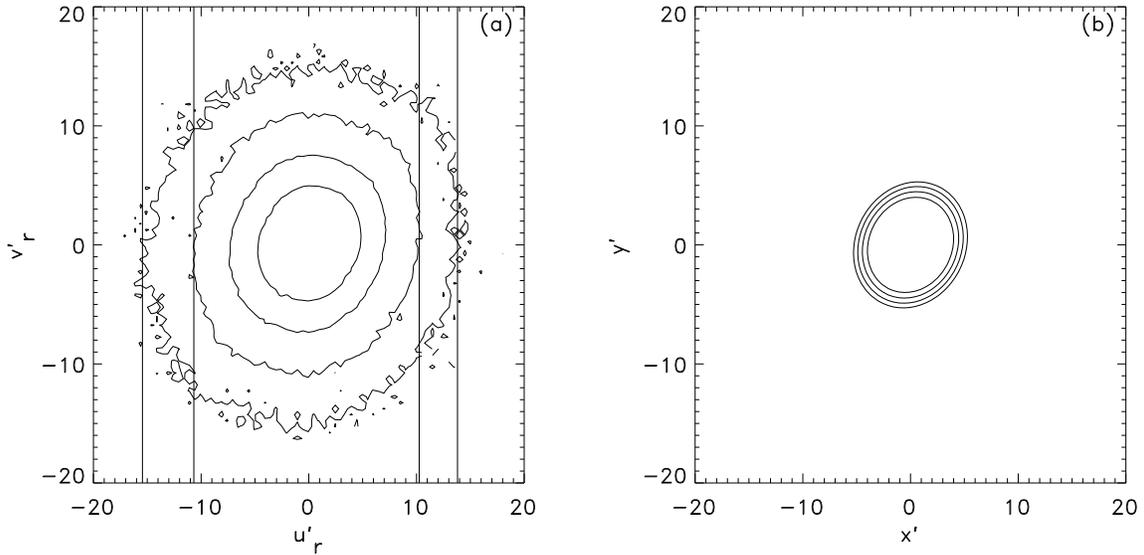}}
  \caption{ Experimental joint PDF, tail part. The indicated levels correspond to 
$e^{-8},~e^{-10},~e^{-12},~e^{-14}$ of the maximum of the PDF, inside out correspondingly.
The contours a) depict experimental PDF, and b) -- Gaussian anisotropic (\ref{Gauss1A}).
}
\vspace{1cm}
\label{fig2}
\end{figure}

 Figures \ref{fig1} and \ref{fig2}
present this PDF in different ranges. We (loosely) define these regions as core part,
Fig. \ref{fig1}, and tail part, Fig. \ref{fig2}. Clearly, Fig. \ref{fig1} corresponds to
the main events, while Fig. \ref{fig2} - to the rare and violent events, -- as indicated by
the levels given in the captures. 

It is clear that the main events, Fig. \ref{fig1}, are not much different from Gaussian.
Both PDF's are noticeably anisotropic, that is the levels are roughly ellipses with  big
axises inclined at some angles to the x-axis. The positions of the levels are roughly the same.
The only difference is asymmetry, the latter being absent in the Gaussian PDF by construction.

The asymmetry of the experimental PDF is evident from the following observations.
 The left edge of the outer level is at $u'_r=-4.1$, while the right
 edge
of it is at $u'_r=3.8$; the left edge of the next level is at $u'_r=-2.5$, while the right
edge is at $u'_r=2.4$. 

Note that both anisotropy and asymmetry are expected to be manifested
by the main events.

Consider now Fig.  \ref{fig2}. We note first that now there is dramatic difference between
the experimental contours and Gaussian. Namely, the experimental contours are much further
away from the core values than the corresponding Gaussian. This feature is however anticipated,
simply corresponding to the presence of tails, that is, to intermittency
 -- in both longitudinal and transverse velocity component
increments. 

On the other hand, other features of the rare events PDF are more surprising.
The experimental contours in the Fig.  \ref{fig2}(a) look roughly similar to those in Fig.  
\ref{fig1}(a), only rescaled to different values, and -- naturally -- more ragged. Indeed, we
see anisotropy -- ellipses with axes inclined roughly the same way in both figures. And, what
is more important, we notice asymmetry in the rare events PDF.  The left edge of the outer 
level corresponds to $u'_r=-15.4$, while the right
 edge
of it is at $u'_r=13.8$; the left edge of the next level corresponds to $u'_r=-10.7$, while the right
edge is at $u'_r=10.2$. 

If we characterize the asymmetry by the ratio of the distance from
the left edge of the level to zero to the distance from the right edge of the level to zero, we
will get for both figures \ref{fig1} and \ref{fig2} -- for the external levels the number $1.1$,
while for the next levels (again for the both figures) the value of $1.05$. Thus, approximately,
the asymmetry is the same in both typical and rare events.

\section{The ${v'}_r^2$ versus ${u'}_r$ distribution: Function $\Phi$, (\ref{Phi})}
\subsection{Global features}
\label{global}
\begin{figure}
  \centerline{\includegraphics[height=8cm,width=16cm]{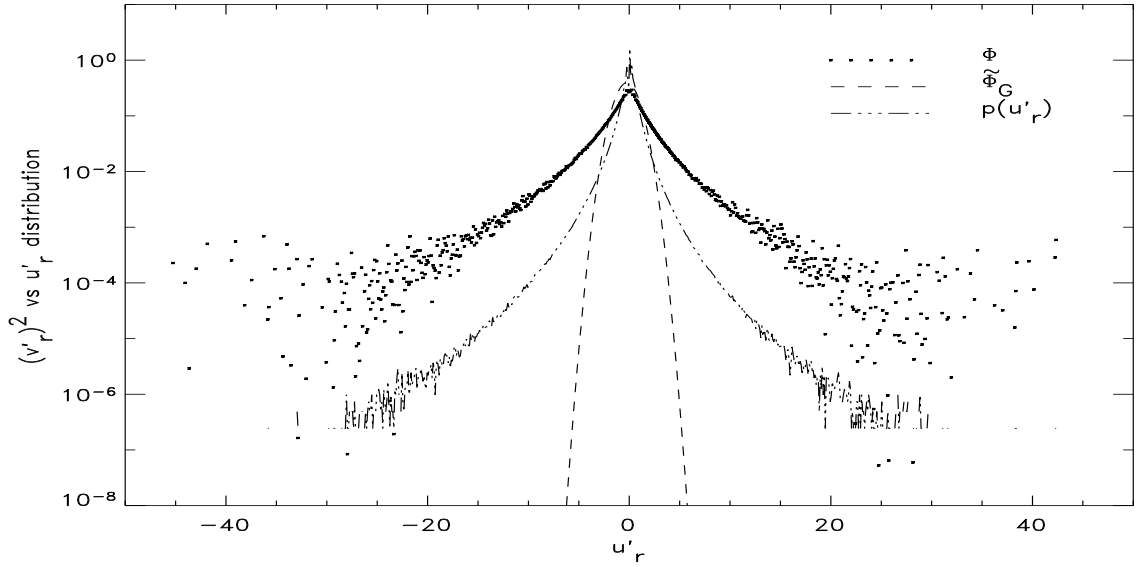}}
  \caption{ Experimental ${v'}_r^2$ distribution, compared with its Gaussian analog, and
with $p({u'}_r)$.
}
\vspace{1cm}
\label{fig3}
\end{figure}

Figure \ref{fig3} presents experimental ${v'}_r^2$ versus ${u'}_r$ distribution. It is compared
with its Gaussian analog on one hand, and with regular experimental distribution $p({u'}_r)$ 
on the other.
All three distributions in the figure are normalized on unity, i.e.,
$$
\int \Phi({u'}_r)d{u'}_r=\int\tilde{\Phi}_G({u'}_r)d{u'}_r=\int p({u'}_r)d{u'}_r=1,
$$
and therefore their direct comparison makes sense. 

It is clear that ${v'}_r^2$-distribution has quite extensive tails. 
 Naturally, the Gaussian distribution is much lower outside
 its core part. Moreover, the tail parts
of $\Phi({u'}_r)$ are much above corresponding parts of $p({u'}_r)$. This is because
the ${v'}_r^2$-distribution is a second moment, see definition (\ref{Phi}), while $p({u'}_r)$
is a zeroth moment of the same distribution,
$$
p({u'}_r)=\int p({u'}_r,{v'}_r)d{v'}_r.
$$
 Note that the
${v'}_r^2$-distribution is much more dispersed at the tail parts as compared with $p({u'}_r)$
in the same areas. Analogous (huge) dispersion of data is observed in conditional average, see below
Fig. \ref{fig7}, and is discussed in Subsection \ref{global_cond}. 
We note here only that this dispersion is attributed to the presence of intermittency.

\subsection{Contribution of the rare violent events}
\label{rare}
Additional information about the tails of the  ${v'}_r^2$ versus ${u'}_r$ 
distribution can be obtained from
 cumulative moments,
\begin{equation}
\langle v_r^2\rangle\rule[-2mm]{.5mm}{6mm}_{|u_r| \ge t}=\int_{-\infty}^{-t} 
\Phi(u_r)du_r+\int_t^\infty \Phi(u_r)du_r,
\label{Phi0t}
\end{equation}
that is, the contribution of events with $|u_r| \ge t$ into the moment $\langle v_r^2\rangle$,
see definition of this moment in (\ref{Phi0}). These moments will be plotted against 
\begin{equation}
P(t)=\int_{-\infty}^{-t} 
p(u_r)du_r+\int_t^\infty p(u_r)du_r,
\end{equation}
 the probabilities of these events.
\begin{figure}
  \centerline{\includegraphics[height=8cm,width=16cm]{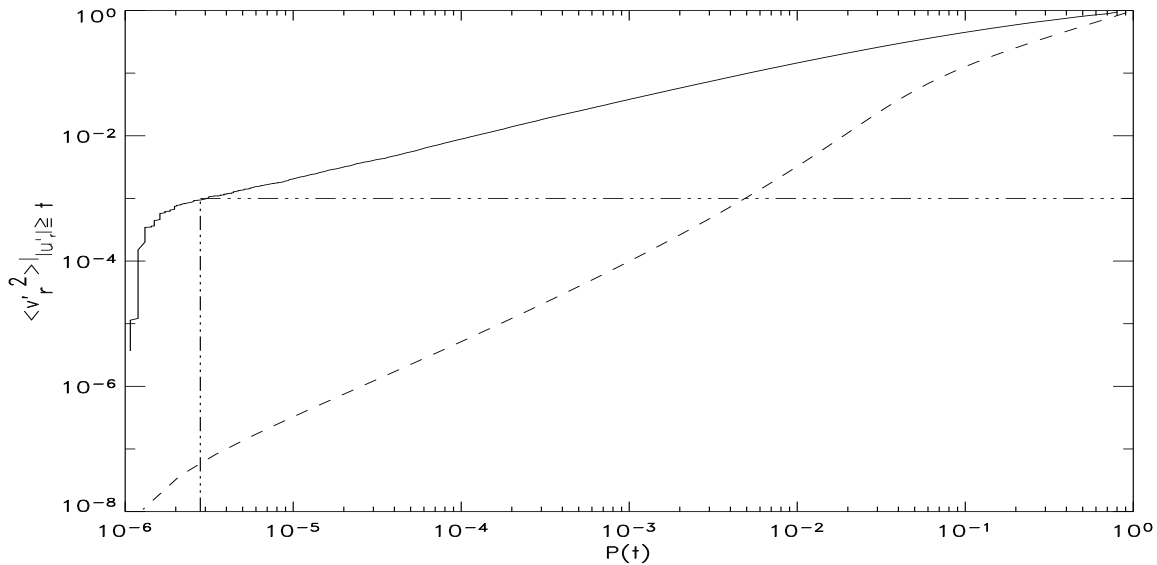}}
  \caption{ Experimental ${v'}_r^2$-distribution for  $|{u'}_r| \ge t$ events versus the
probability of these events $P(t)$ (solid line), compared with its Gaussian counterpart (dashed
line).
}
\vspace{1cm}
\label{fig3a}
\end{figure}

 Figure \ref{fig3a} presents such a plot. The experimental distribution is compared with
its Gaussian counterpart. For $t=0$, all events are presented, and therefore $P(t=0)=1$ and
$\langle {v'}_r^2\rangle\rule[-2mm]{.5mm}{6mm}_{|{u'}_r| \ge 0}=1$. If, on the other hand, 
$t\to\infty$, then both distributions go to zero. 

The difference between experimental and
Gaussian cases is quite substantial. As an example (depicted by dashed-dotted straight lines),
 we see  that 
$\langle {v'}_r^2\rangle$ reaches $10^{-3}$ fraction of its final value with only $2.8\times 
10^{-6}$ part of all events. According to our estimate, these events correspond to quite
violent outbursts with $t\ge 27.4$ As seen from Fig. \ref{fig3}, these events are almost at 
the very end of the measured tails. 
 The Gaussian counterpart reaches the same value of $10^{-3}$
with $4.7\times 10^{-3}$ part of events; both these  numbers are comparable for 
the Gaussian distribution. 

On the other hand, the $\langle {v'}_r^2\rangle$  of the Gaussian counterpart with
the same probability $2.8\times 10^{-6}$ as the experimental reaches only 
$5.8\times 10^{-8}$ fraction of
its final value, these numbers being again more or less comparable. 

\subsection{Asymmetry}
\begin{figure}
  \centerline{\includegraphics[height=8cm,width=16cm]{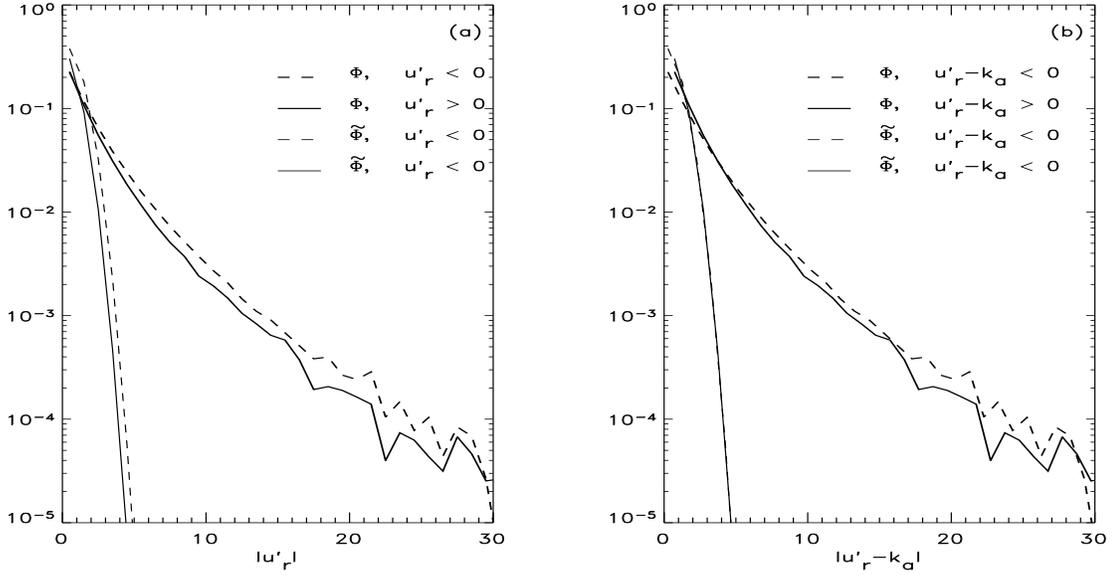}}
  \caption{ Experimental ${v'}_r^2$-distribution. Direct comparison of the left and right
wings.
}
\vspace{1cm}
\label{fig4}
\end{figure}

Note that all three distributions depicted in Fig. \ref{fig3} are constructed with the same 
resolution, that is, the bin-size for all three distribution was the same and equal to $0.1$.
This makes it possible to compare the dispersion of the data for different 
distributions. On the other hand,  huge 
dispersion of the ${v'}_r^2$-distribution does not make it possible to compare its tails. In
order to do this, we have to have a more smooth PDF, and that can be done by constructing the
distribution with a larger bin-size. Figure \ref{fig4} presents the ${v'}_r^2$-distribution
constructed with a bin-size $=1.$

Figure \ref{fig4}(a) compares the positive and negative parts of the distribution directly. It
is obvious that the negative tail prevails over the positive for rather
big ${u'}_r$. That is, the rare
stormy events are definitely asymmetric. 

The ${v'}_r^2$-distribution can be considered as a PDF
centered at ${u'}_r=k_a$, see  (\ref{mixed2}). That is, the distribution is definitely
asymmetric, and theoretically, this asymmetry {\it should be} present at the tails as well. 
This actually 
can be seen from the shifted Gaussian distribution $\tilde{\Phi}$, 
see definition (\ref{GaussPhiA}) and (\ref{GaussPhi1}), and
 from the Fig. \ref{fig4}(a), that the negative part of the distribution
is above the positive for any ${u'}_r$. 

In order to see if the ${v'}_r^2$-distribution is just a shifted distribution, or not, we make
direct comparison of
its part where ${u'}_r-k_a$ is positive with the part where ${u'}_r-k_a$ is negative, see 
\ref{fig4}(b). The two parts of the corresponding Gaussian distribution $\tilde{\Phi}$, of course, coincide
(and cannot be distinguished in the plot),
while the experimental distribution still shows asymmetry: The centered negative part still prevails.
This confirms the above conclusion that the rare violent events are asymmetric.

Another test for asymmetry is
  measuring the contribution of rare events, as in Subsection \ref{rare}.
Namely, we will consider cumulative moments,
\begin{equation}
\langle u_r v_r^2\rangle\rule[-2mm]{.5mm}{6mm}_{|u_r| \ge t}=\int_{-\infty}^{-t} 
u_r\Phi(u_r)du_r+\int_t^\infty u_r\Phi(u_r)du_r,
\label{kat}
\end{equation}
cf. (\ref{Phi0t}). Or, in dimensionless form,
\begin{equation}
k_a(t)=\langle {u'}_r {v'}_r^2\rangle\rule[-2mm]{.5mm}{6mm}_{|{u'}_r| \ge t}=\int_{-\infty}^{-t} 
{u'}_r\Phi({u'}_r)d{u'}_r+\int_t^\infty {u'}_r\Phi({u'}_r)d{u'}_r,
\label{kat1}
\end{equation}
\begin{figure}
  \centerline{\includegraphics[height=8cm,width=16cm]{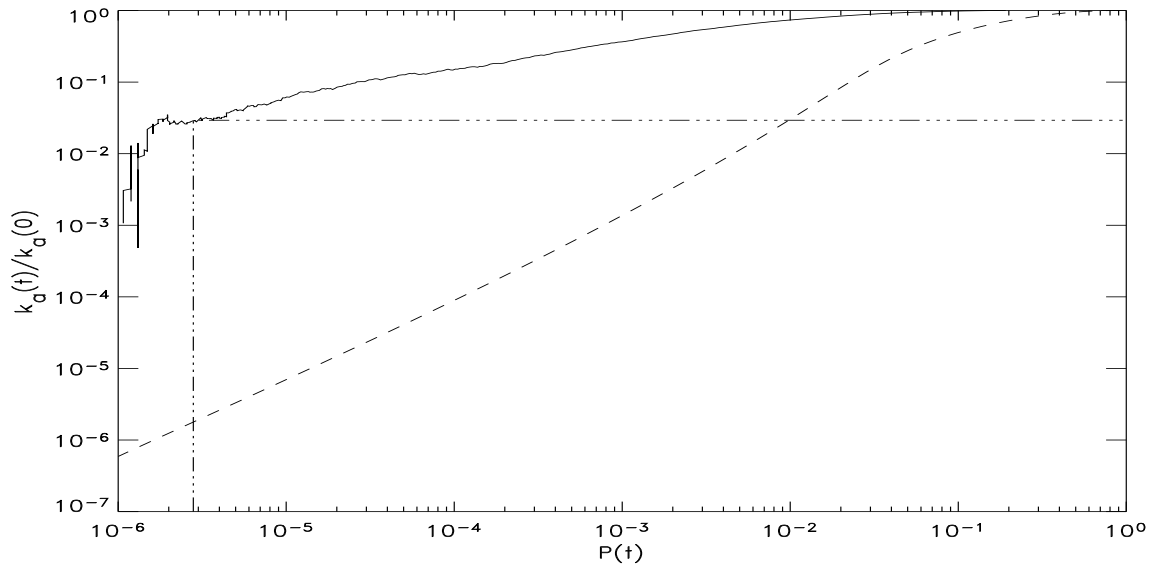}}
  \caption{ Experimental $k_a(t)$ (defined in (\ref{kat1})) for $|{u'}_r| \ge t$ events versus
the probability of these events $P(t)$. 
}
\vspace{1cm}
\label{fig6}
\end{figure}
The case $t=0$ corresponds to all events, so that $k_a(t=0)$ assumes its final value, and at $t>0$ it 
approaches
it. The case $t\to\infty$ corresponds to $k_a=0$. Figure \ref{fig6} illustrates $k_a(t)/k_a(0)$, or
actually it shows how the the third moment (\ref{mixed2}) is formed.

Obviously, there is a big difference between the experimental $k_a(t)/k_a(0)$ and its Gaussian
counterpart, as seen from Fig. \ref{fig6}. As an example (see the dashed-dotted 
straight lines in the figure), we took the same probability $2.8\times
10^{-6}$ as in Fig. \ref{fig3a}, corresponding to extremely violent events. This time, it
reaches $0.03$th fraction of its final value. In contrast, its Gaussian counterpart reaches the
same fraction with $0.01$ part of events, -- these two numbers ($0.03$ and
$0.01$) being comparable (cf. Subsection \ref{rare}).
On the other hand, the Gaussian counterpart with probability $2.8\times 10^{-6}$ reaches the value of
$1.8\times 10^{-6}$. Again, these two numbers are comparable.

Thus, the contribution of the rare violent events into the experimental moment (\ref{mixed2}) is
substantial, as opposed to the "regular" situation presumably illustrated by Gaussian distribution.
Recall that this (odd) moment does not vanish due to asymmetry of the distributions, and therefore
this substantial contribution of the rare violent events into $k_a$-moment means that these
events also possess asymmetry.

\section{Conditional averages}
\subsection{Global properties}
\label{global_cond}
 \begin{figure}
  \centerline{\includegraphics[height=8cm,width=16cm]{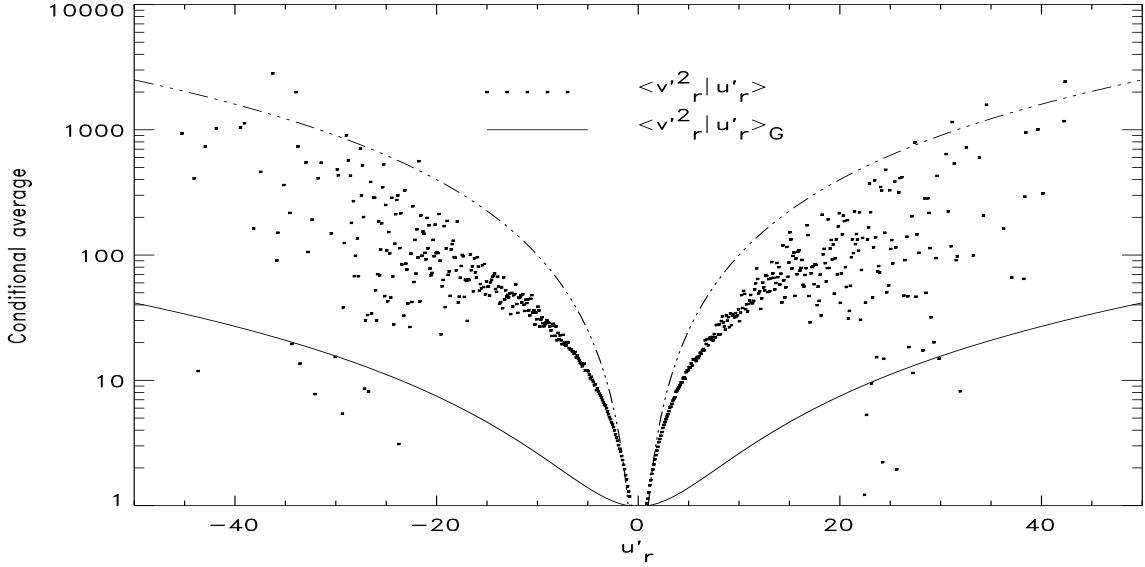}}
  \caption{ Experimental conditional average $\langle {v'}_r^2|{u'}_r\rangle$, compared
with Gaussian defined in (\ref{cond1}), and with "ideal" correlation, depicted by 
dashed-dotted line.
}
\vspace{1cm}
\label{fig7}
\end{figure}
It is interesting to note that the the experimental conditional average 
$\langle {v'}_r^2|{u'}_r\rangle$ is much above the
Gaussian defined in (\ref{cond1}), see Fig. \ref{fig7}.
 If these two variables, ${u'}_r$ and ${v'}_r$ would
be statistically independent, then the conditional average is unity, well below
experimental average. The anisotropic Gaussian distribution (for which ${u'}_r$ and ${v'}_r$
variables are related, and therefore so are  ${u'}_r$ and ${v'}_r^2$), is still much lower
than experimental, as seen from the figure.

In another extreme case (as opposed to statistically independent variables) we have an  "ideal"
correlation,  ${v'}_r=\alpha{u'}_r$, where $\alpha$ is a constant, which is, due to our
normalization, equal to unity. Then, simply,  
$$\langle {v'}_r^2|{u'}_r\rangle={u'}_r^2$$
(cf. with Gaussian conditional average with maximal correlation coefficient $C=\pm 1$, from
(\ref{cond1})). As seen from the Fig. \ref{fig7}, this conditional average is even higher than
the experimental value: Of course, any connection between two variables is less than "ideal".

Another remarkable feature of this average is its gigantic dispersion, 
 cf. Subsection \ref{global}. Perhaps, the simplest way to explain
it is to consider the two variables ${u'}_r$ and ${v'}_r$ statistically independent.
As mentioned above, in this case, theoretically $\langle {v'}_r^2|{u'}_r\rangle=1$.
However, experimental measurements would give dispersed values, the statistics being defined
by the number of events. To be more specific, for small and moderate values of $|{u'}_r|$
with huge number of events, the data of $\langle {v}_r^2\rangle$ would be quite close 
to unity. For big values of
$|{u'}_r|$, with only few events the data would be strongly dispersed around unity. 
In particular,
if there is only one event in some bins, then the values of $\langle {v}_r^2\rangle$ would
coincide with  $ {v}_r^2$ themselves, and they would be quite dispersed if the process is intermittent. 
Qualitatively, the experimental conditional average in the Fig. \ref{fig7} looks as
described above, that is, it is smooth for small and moderate values of  $|{u'}_r|$,
and strongly dispersed for big values. Returning to ${v'}_r^2$ vs ${u'}_r$ distribution, i.e.,
function $\Phi({u'}_r)$, we recall that $\Phi({u'}_r)=\langle {v'}_r^2|{u'}_r\rangle p({u'}_r)$,
see (\ref{mixed_c_d}),
where, $p({u'}_r)$ is not very dispersed function, as seen from Fig. \ref{fig3}. Therefore,
the dispersion of the ${v'}_r^2$ vs ${u'}_r$ distribution is explained analogously to
the dispersion of conditional average $\langle {v'}_r^2|{u'}_r\rangle$.

\subsection{Asymmetry}
\begin{figure}
  \centerline{\includegraphics[height=8cm,width=16cm]{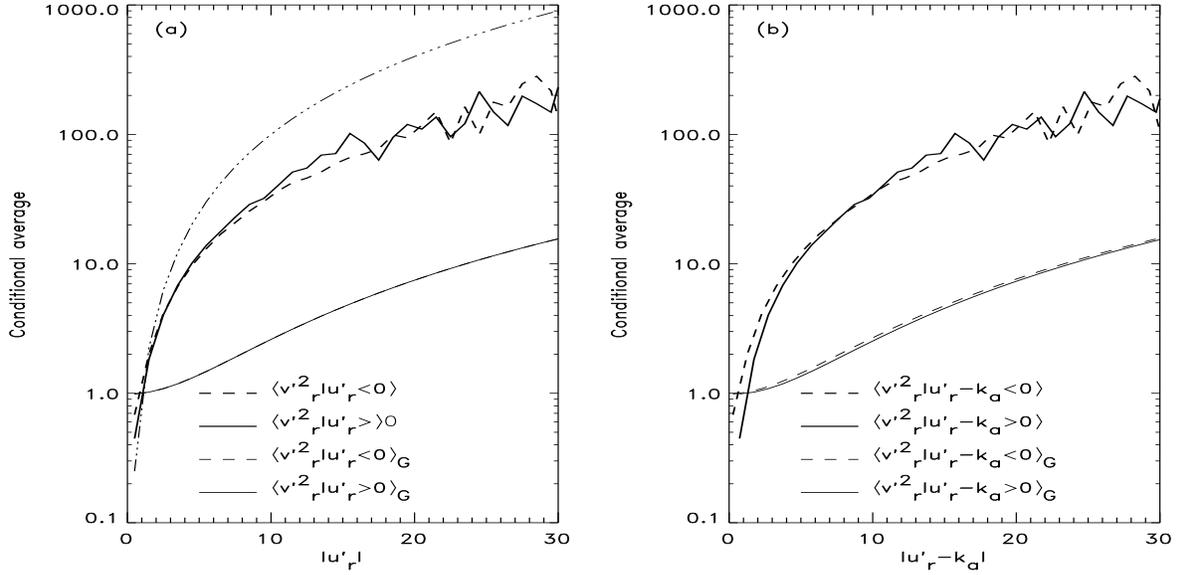}}
  \caption{ Direct comparison of negative and positive parts of conditional average
$\langle {v'}_r^2|{u'}_r\rangle$. Dashed-dotted line corresponds to "ideal" correlation.
}
\vspace{1cm}
\label{fig8}
\end{figure}
The  data in Fig. \ref{fig7} are presented with relatively high resolution, the bin-size being
equal to $0.1$. This makes it possible to see the dispersion and to interpret it as in 
Subsection \ref{global_cond}. On the other hand, it is difficult to study the asymmetry with
these dispersed data. Another processing of data with bin-size equal to $1$ are presented
in Fig. \ref{fig8}, where we compare the negative and positive parts of this conditional average
directly.

We note here that Gaussian expression for conditional average (\ref{cond1}) calculated from
the joint 2D distribution (\ref{Gauss1A}) and used in Fig. 
\ref{fig8} is symmetric (recall that it is not known how to include asymmetry into Gaussian
or some other simple test 2D distribution). For this reason, the difference between the
right and left Gaussian wings in both figures (a) and (b) is spurious. Still, putting the Gaussian
conditional average into these two plots seems to be useful in order to  compare
the experimental data with something "regular".

Both quantities, $\langle {v'}_r^2\rangle$ vs ${u'}_r$ distribution, Fig. \ref{fig4}, and 
$\langle {v'}_r^2|{u'}_r\rangle$,
Fig.  \ref{fig8}, are related. For that reason in the latter figure we also check the symmetry
in respect to the ${u'}_r=k_a$ point (in \ref{fig8}(b), analogously to \ref{fig4}(b)), in addition
to the symmetry in respect to ${u'}_r=0$ point, given in \ref{fig4}(a) and \ref{fig8}(a). In other
words, we check if the conditional average $\langle {v'}_r^2|{u'}_r\rangle$ is simply a shifted 
(but still symmetric) distribution, or not.

Close examination of Fig. \ref{fig8} suggests that there is no substantial asymmetry 
{\it as a systematic trend} in this conditional average.  Although this is an unexpected 
conclusion, there might be an explanation as follows. 
As can be seen from Fig. \ref{fig4}, the $\langle{v'}_r^2\rangle$ distribution, i.e., the
function $\Phi({u'}_r)$, is asymmetric, and so is the ${u'}_r$ distribution, $p({u'}_r)$, see
\cite{asym}. This
means that, in particular, the negative wings of both quantities are
 elevated above the right-hand wings. According
to (\ref{mixed_c_d}), the conditional average  $\langle{v'}_r^2|{u'}_r\rangle$ is defined as
a ratio of these two quantities. This suggests that increased value of  $\Phi({u'}_r)$ in
numerator is balanced by increased value of $p({u'}_r)$ in denominator, decreasing the asymmetry
of $\langle{v'}_r^2|{u'}_r\rangle$. In other words, the latter quantity appears to be less
sensitive to the asymmetry than the former two.
 
\section{Discussion}

The asymmetry of turbulence has been studied for a long time. Its relation to the vorticity
production was pointed out 
by \cite{Betchov}.
The asymmetry of $u_r$-distribution resulting in the 4/5-Kolmogorov law was interpreted with
the help of the ramp-model, see  \cite{94}, \cite{96a}: As  $\langle u_r\rangle=0$, any compression,
 with $u_r<0$, is as efficient as expansion, with $u_r>0$. However, the compression appears 
stronger but rarer
than decompression, the latter being weaker and longer. Therefore, $\langle u_r^3\rangle\not= 0$.
This model immediately suggests that this asymmetry is related to the intermittency: The
compressed rare but strong events are supposed to be intermittent.

The ramp-model is only heuristic, however. It was shown by \cite{1}, \cite{2}, \cite{111}
  that Burgers vortex, 
embedded into a converging motion, acquires negative skewness, this picture containing both
asymmetry and intermittency. As the ramp-model does not exactly correspond to this picture,
it had to be modified. This modification was called the bump-model, see \cite{jfm}. Actually, 
``the bump" corresponds to the Burgers vortex. This model is now two-dimensional, -- to
include transverse velocity increments $v_r$ into the picture. In particular, it 
automatically explains why the
 mixed third moment $\langle u_rv_r^2\rangle$,  (\ref{mixed0}), does not vanish. 

Fully two-dimensional PDF describing all three items, namely asymmetry, anisotropy and 
intermittency,
was discussed in this paper. The PDF, and also some additional conditional and cumulative averages
seem to support the idea about the connection between these three features. In the whole picture, 
there is
additional (to the Kolmogorov cascade) energy transport from the large non-universal scales
directly to the small scales. We note that the Burgers vortex, with a small radius, is generated
by a relatively large-scale motion, thus making it possible to transfer the energy directly from
large scales to small. In the framework of the bump-model, it is ``the bump" which is generated
analogously to the Burgers vortex. The scale of the bump is definitely much smaller than the
scale of the generating it motion (which can be seen from Fig. 5(c) by \cite{jfm}), and that could
explain the direct interaction between the non-universal large scales and small scales.

I thank K. R. Sreenivasan and  S. Kurien for the data and for useful discussions.

\newpage
\appendix{\bf Appendix}

The Gaussian joint PDF has the form,
\begin{equation}
p_G(u_r,v_r)=\frac{1}{2\pi\sqrt{\langle u_r^2\rangle\langle v_r^2\rangle-
\langle u_rv_r\rangle^2}}
\exp{\left\{-\frac{u_r^2\langle v_r^2\rangle-2u_rv_r\langle u_rv_r\rangle+v_r^2\langle u_r^2\rangle}
{2(\langle u_r^2\rangle\langle v_r^2\rangle-
\langle u_rv_r\rangle^2)}\right\}},
\label{GaussA}
\end{equation}
$\langle u_r\rangle=\langle v_r\rangle=0$.
Or, for dimensionless variables,

\begin{equation}
p_G({u'}_r,{v'}_r)=\frac{1}{2\pi\sqrt{1-C^2}}\exp{\left\{ -\frac{{u'}_r^2-2{u'}_r{v'}_rC+{v'}_r^2}
{2(1-C^2)} \right\}}
\label{Gauss1A}
\end{equation}
where
\begin{equation}
C=\langle {u'}_r {v'}_r\rangle=\frac{\langle u_rv_r\rangle}{\langle u_r^2\rangle^{1/2}
\langle v_r^2\rangle^{1/2}},
\end{equation}
the correlation coefficient. Note that in isotropic turbulence $C$ vanishes, and therefore
the distribution (\ref{GaussA}), or (\ref{Gauss1A}), corresponds to anisotropic Gaussian 
process. Note, however, that this distribution is not asymmetric. Indeed, it is easy to
show that
$\langle u_r^3\rangle_G =\int u_r^3p_G(u_r,v_r)du_rdv_r=0$ and
$\langle u_rv_r^2\rangle_G =\int u_rv_r^2p_G(u_r,v_r)du_rdv_r=0$.

For Gaussian distribution (\ref{GaussA}), 
\begin{equation}
\Phi_G(u_r)=\frac{1}{\sqrt{2\pi}}\left[\frac{\langle u_r^2\rangle\langle v_r^2\rangle-
\langle u_rv_r\rangle^2}{\langle u_r^2\rangle^{3/2}} 
+u_r^2\frac{\langle u_rv_r \rangle^2}{\langle u_r^2\rangle^{5/2}}\right]
\exp{\left\{-\frac{u_r^2}{2\langle u_r^2\rangle}\right\}}
\label{GaussPhiA}
\end{equation}

Or, for dimensionless variables, we have,
\begin{equation}
\Phi_G({u'}_r)=\frac{1}{\sqrt{2\pi}}[1-C^2+{u'}_r^2C^2]\exp{\left\{-\frac{{u'}_r^2}{2}
\right\}}
\end{equation}

For Gaussian distribution, according to (\ref{mixed_c}) and (\ref{GaussPhiA}), we have
\begin{equation}
\langle v_r^2|u_r\rangle_G=\langle v_r^2\rangle 
\left[ 1-C^2+\frac{u_r^2}{\langle u_r^2 \rangle}C^2\right],
\label{cond}
\end{equation}
or
\begin{equation}
\langle {v'}_r^2|u'_r\rangle_G=1-C^2+{u'}_r^2C^2
\label{cond1}
\end{equation}
These Gaussian expressions are useful to check in two limiting cases.
If $C=0$, the two variables $u_r$ and $v_r$ become statistically 
independent, and, naturally, $\langle v_r^2|u_r\rangle=\langle v_r^2\rangle$.
In another limiting case of perfect correlation, $v_r=\alpha u_r$, and
therefore $C=\pm 1$, then
$\langle v_r^2|u_r\rangle=\alpha^2 u_r^2=\langle v_r^2\rangle u_r^2/\langle u_r^2\rangle$. These
two extreme cases are indeed confirmed by (\ref{cond}).

As mentioned above, the Gaussian distribution (\ref{GaussA}) is not asymmetric.
It is not known how to construct a simple
test 2D PDF containing both anisotropy and asymmetry. We therefore are going to use
(\ref{GaussA}) directly comparing it with the experimental joint 2D PDF. However, in all
expressions containing 1D PDF, instead of symmetric 1D PDF resulting from integration of the
PDF (\ref{Gauss1A}) over $dv'_r$, we will use a simple asymmetric
PDF, $I_r$, constructed as a sum of two Gaussian distributions,  see \cite{asym} and \cite{jfm}, 
so that
$$\int I_rdu'_r=1,~~ \int u'_rI_rdu'_r=\langle u'_r\rangle=0,~~  
\int {u'_r}^2I_rdu'_r=\langle {u'_r}^2\rangle=1,~~ {\rm and}~~\int {u'_r}^3
I_rdu'_r=\langle {u'_r}^3\rangle.$$  
Being asymmetric it is otherwise almost indistinguishable from Gaussian 
1D PDF following from (\ref{Gauss1A}), at least visually.

The Gaussian $\Phi_G(x)$ should be  "corrected" as well  to incorporate asymmetry. We will
therefore use 
\begin{equation}
\tilde{\Phi}_G(x')=\Phi_G(x'-k_a),
\label{GaussPhi1}
\end{equation} 
instead of $\Phi_G(x)$ defined in (\ref{GaussPhiA}). Then, of course,
$$
\langle {u'}_r{v'}_r^2\rangle_G=\int {u'}_r\tilde{\Phi}_G({u'}_r)d{u'}_r=k_a.
$$
Note that the difference between the plots of $\Phi_G$ and $\tilde\Phi_G$ is only cosmetic.

\newpage

\bibliography{anis}

\end{document}